# Ultrabroadband nanocavity of hyperbolic phonon polaritons in 1D-like α-MoO₃

Ingrid D. Barcelos[1†\*], Thalita A. Canassa[2\*], Rafael A. Mayer[1,3\*†], Flavio H. Feres[1,3], Eynara G. de Oliveira[4], Alem-Mar B. Goncalves[2,4], Hans A. Bechtel[5], Raul O. Freitas[1], Francisco C. B. Maia[1], Diego C. B. Alves[2,4]

[1] *Brazilian Synchrotron Light Laboratory (LNLS), Brazilian Center for Research in Energy and Materials (CNPEM), Zip Code 13083-970, Campinas, Sao Paulo, Brazil*.

[2] *Instituto de Química, Universidade Federal de Mato Grosso do Sul (UFMS), Zip Code 79070-900, Campo Grande, Mato Grosso do Sul, Brazil.*

[3] *Physics Department, Gleb Wataghin Physics Institute, University of Campinas (Unicamp), 13083-859 Campinas, Sao Paulo, Brazil*

[4] *Instituto de Física, Universidade Federal de Mato Grosso do Sul (UFMS), Zip Code 79070-900, Campo Grande, Mato Grosso do Sul, Brazil.*

[5] *Advanced Light Source, Lawrence Berkeley National Laboratory, Berkeley, California 94720, USA.*

\*These authors contributed equally to this work.
†Corresponding-Author: ingrid.barcelos@lnls.br, rafael.mayer@lnls.br

Abstract

The exploitation of phonon-polaritons in nanostructured materials offers a pathway to manipulate infrared (IR) light for nanophotonic applications. Notably, hyperbolic phonons polaritons (HP$^2$) in polar bidimensional crystals have been used to demonstrate strong electromagnetic field confinement, ultraslow group velocities, and long lifetimes (~ up to 8 ps). Here we present nanobelts of α-phase molybdenum trioxide (α-MoO$_3$) as a low-dimensional medium supporting HP$^2$ modes in the mid- and far-IR ranges. By real-space nanoimaging, with IR illuminations provided by synchrotron and tunable lasers, we observe that such HP$^2$ response happens via formation of Fabry-Perot resonances. We remark an anisotropic propagation which critically depends on the frequency range. Our findings are supported by the convergence of experiment, theory, and numerical simulations. Our work shows that the low dimensionality of natural nanostructured crystals, like α-MoO$_3$ nanobelts, provides an attractive platform to study polaritonic light-matter interactions and offer appealing cavity properties that could be harnessed in future designs of compact nanophotonic devices.

**Introduction**

MoO$_3$ has received considerable attention in a wide range of applications including inorganic light-emitting diodes[1], Li-ion batteries[2–4], electrochemical supercapacitors[5,6], electrochromic devices[7] and others[8,9]. The α-phase of this material (α-MoO$_3$) has been synthesized in a variety of nanostructures: nanowires[10], nanobelts[11,12], nanorods[13], two-dimensional (2D) crystals[14], among others[15]. In particular, α-MoO$_3$, as a typical van der Waals (vdW) semiconductor, has shown an exquisite optical response[16,17] caused by its crystalline anisotropy. This crystal has an orthorhombic layered oxide structure[18,19] composed of distorted and edge-shared MoO$_6$ octahedra, wherein each octahedron contains three types of oxygen sites: the terminal O$_1$, which is doubly bonded with the Mo atom, the doubly coordinated and asymmetric bridging O$_2$, and the triply coordinated and symmetric bridging O$_3$, as shown in Figure 1a. The α-MoO$_3$ structural anisotropy can be described by a layered structure parallel to the (010) plane[20,21]. Each layer is then composed of two sub-layers, each of which formed by corner-sharing octahedra along the [001] and [100] planes (Figure 1b).

Because of this atomic configuration, α-MoO$_3$ is optically biaxial, and in 2D shapes exhibits directional hyperbolic phonon-polariton (HP$^2$) waves[17,20–22] whose photonic properties depend on the crystal thickness and structural anisotropy. Fundamentally, HP$^2$ modes[23–26] are quasi-particles formed by the coupling of electromagnetic field excitation to lattice vibrations of a polar medium with an anisotropic and phononic resonant electrical permittivity $\overleftrightarrow{\varepsilon} = (\varepsilon_{xx}, \varepsilon_{yy}, \varepsilon_{zz})$, where $\overleftrightarrow{\varepsilon}$ is a diagonal second-rank tensor valid for the α-MoO$_3$ case. A hyperbolic medium is defined within spectral regions where the real part of two components of the $\overleftrightarrow{\varepsilon}$ have opposite sign. In general, this condition occurs inside a 'Reststrahlen band (RB)', which is the spectral window delimited between the transverse $\omega_{TO}$ and longitudinal $\omega_{LO}$ optical phonons[26]. Therefore, the HP² waves are characterized by modes of the frequencies ($\omega$) inside the RBs of a hyperbolic medium. Forming open hyperboloidal-shaped isofrequency surfaces of type I ($\varepsilon_{zz} < 0$, but $\varepsilon_{xx} > 0$ and/or $\varepsilon_{yy} > 0$) or type II ($\varepsilon_{xx} < 0$ and/or $\varepsilon_{yy} < 0$, but $\varepsilon_{zz} > 0$) in the momenta space, the HP² possess high momentum ($q$). Their photonic properties is, typically, described in terms of the corresponding type I or II $\omega - q$ dispersion relations[27]. In the case of 2D crystals of α-MoO₃, strong phonons[18] enable HP² waves[17,20,21,28,29] in an ultra-broadband frequency range spanning from 250 cm⁻¹, in the THz domain, to 1200 cm⁻¹ in the mid-IR.

In this work we exploit the polaritonic response of α-MoO₃ nanobelts, with reduced dimensionality approaching "1D-like" (Figure 1c), using synchrotron infrared nano-spectroscopy (SINS)[30–33] and scattering-scanning near-field optical microscopy (s-SNOM). As reported[34], such nanoscopy techniques (see details in supplementary material) possess a high spatial resolution and momentum nanoprobe allowing for full characterization of HP² waves. The polaritons in α-MoO₃ nanobelts feature Fabry-Perot (FP) resonances, primarily, from type I HP²s[20,22] in the 970-1100 cm⁻¹ range. The modes show dependence on the geometric form factor (thickness and width) of the nanobelt and propagate along the [100] crystal direction. In the same crystalline axis, we also show the first infrared nanoimages of type I polaritonic modes in the far-IR band extending from 440 to 490 cm⁻¹. Those analyses are confirmed by a robust convergence among experiment and numerical simulations. Hence, α-MoO₃ nanobelts are presented as an HP² nanocavity directly synthesized without the need of lithography or other post-synthetic sculpting processes. This low-dimensional polaritonic material opens new avenues for integrated planar IR photonics[35], dictating polaritons control[36] and nanoresonators for phonon-enhanced molecular vibrational spectroscopy[37].

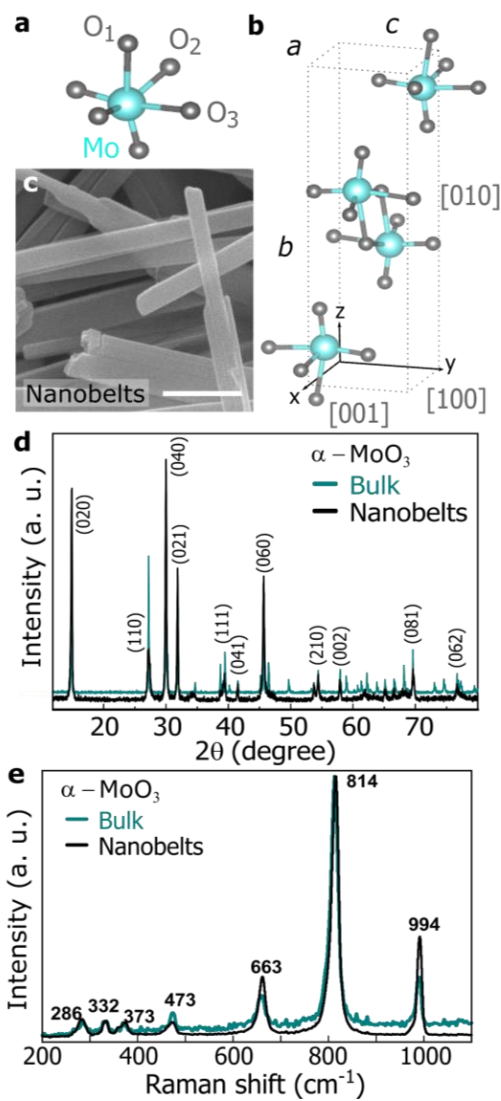

**Figure 1| Overview of α-MoO₃ nanobelts crystal structure and morphology.** (a) Scheme of the three oxygen sites of asymmetric Mo–O bonds along the different crystalline axes labeled as $O_1$, $O_2$, and $O_3$ of α-MoO₃. (b) The unit cell of α-MoO₃ shown by the dashed lines with the lattice constants: $a$ = 0.396 nm, $b$ = 1.385 nm and $c$ = 0.369 nm. Green/gray spheres represent molybdenum/oxygen atoms. The principal axes of the crystal, x, y, and z coincide with the crystalline directions [100], [001], and [010], respectively. (c) SEM images of α-MoO₃ nanobelts (scale bar: 1 mm) on the substrate. (d) Corresponding XRD patterns: nanobelts (black) and bulk (green). (e) Raman spectra of α-MoO₃ nanobelts and bulk crystal on a glass substrate.

**Results**

The α-MoO$_3$ nanobelts were synthesized by the hydrothermal method[12] that yields an orthorhombic crystalline structure in which all three lattice parameters (*a*, *b* and *c*) are different as shown in Figure 1b. Representative scanning electron microscopy (SEM) images of the α-MoO$_3$ nanobelts (Figure 1c) reveal well-defined rectangular-like shapes for the synthesized crystals in which the long-axis corresponds to the [001] direction, as already reported[38–41]. The synthesis produced nanobelts of widths (w) varying between 150 and 400 nm and lengths, from 10 to 30 mm. X-ray diffraction (XRD) reveals the predominance of the α-phase as confirmed by the good correspondence with MoO$_3$ bulk crystals (Figure 1d). The diffraction peaks of the XRD pattern for both species can be readily indexed to be orthorhombic with lattice parameters *a* = 0.396 nm, *b* = 1.385 nm and *c* = 0.369 nm according to the JCPDS database number 05-0508[39,42]. The more intense peaks are identified in the family planes (0 k 0), with k = 2, 4 and 6, which suggests an anisotropic structure. It is well known that α-MoO$_3$ has a lamellar structure along the [010] axis, formed by layers of MoO$_6$ octahedra with preferential growth along the [001] axis[42]. No peaks of other phases were detected, indicating a high crystalline purity of the nanobelts.

The α-MoO$_3$ orthorhombic type has three oxygen species (O$_{1-3}$) along different crystallographic directions giving rise to the various Mo–O bonding types (Figure 1a). Such a complex structure leads to a great variety of phonon modes in the mid- to far-IR as depicted by the nearly coincident Raman spectra of the bulk crystal and nanobelts in Figure 1e. The peaks at 994, 814, 663, 473, 373, 332, and 286 cm$^{-1}$ are assigned to vibrational modes of the orthorhombic α-MoO$_3$ phase[43,44],[18]. The 286 cm$^{-1}$ peak is attributed to O=Mo=O wagging vibrations. The peaks at 332 and 373 cm$^{-1}$ are assigned to O$_3$-Mo-O$_3$ bending mode and O-Mo-O scissoring, respectively. The peak at 473 cm$^{-1}$ corresponds to the Mo–O–Mo stretching and bending vibration. The peak at 663 cm$^{-1}$ stems from triple-coordinated oxygen stretching mode and the peak at 814

cm$^{-1}$, from the O–Mo–O stretching along the [100] direction. The peak at 994 cm$^{-1}$ is due to the oxygen stretching vibration of Mo–O$_1$ bond along the [010] direction.

Four hyperbolic RBs are predicted for α-MoO$_3$ crystals according to the plot of $Re(\varepsilon_{xx})$, $Re(\varepsilon_{yy})$ and $Re(\varepsilon_{zz})$ in Figure 2a. These permittivity components are expressed by the Lorentz model using input parameters from previous reports[18,20,28],37. As illustrated, in a wide IR range, the signs of permittivities along three directions can achieve many different combinations, showing different types and directions of hyperbolicity. Specifically, in RB$_1$ and RB$_4$, where ε$_{xx}$; ε$_{yy}$ > 0 (ε$_{xx}$≠ε$_{yy}$) and ε$_{zz}$ < 0, the out-of-plane dispersion is hyperbolic while the in-plane dispersion is elliptical, and the modes propagate anisotropically in the x-y plane. In RB$_2$, ε$_{yy}$ < 0 and ε$_{xx}$ ; ε$_{zz}$ > 0 (ε$_{xx}$≠ε$_{zz}$), and in RB$_3$, ε$_{xx}$ < 0 and ε$_{yy}$ ; ε$_{zz}$ > 0 (ε$_{yy}$≠ε$_{zz}$), the polaritonic modes show in-plane and out-of-plane hyperbolic dispersion with directional propagating depending on the RB.

Accordingly, such anisotropy determines different field distributions and polariton propagation in each RB as shown by the simulations in Figure 2b. The simulations employed finite-difference time-domain (FDTD) numerical method, where a dipole source was used to excite the polaritons in a semi-infinite nanobelt of w=1 μm, and t=100 nm. The polariton response was obtained by monitoring the real value of the electric field in the z-direction (Re(E$_z$)) on a plane above the crystal surface (see Methods for more details). As one may observe, the polaritons fringes are approximately isotropically distributed in RB$_1$ and RB$_4$, while in RB$_2$ and RB$_3$ the propagation is restricted to [001] and [100] directions, respectively. Thus, the simulations confirm the in-plane elliptical dispersion in RB$_1$ and RB$_4$ and demonstrates the extreme in-plane anisotropy in RB$_2$ and RB$_3$.

The in-plane anisotropy ($\varepsilon_{xx} \neq \varepsilon_{yy}$) of the α-MoO$_3$ crystal leads to different $\omega - q$'s for the [100] and [001] crystalline directions as revealed from the poles of the reflectivity coefficient $r_p(\omega, q)$, visualized by the false color plot of the imaginary part of $r_p(\omega, q)$ in Figure 2c for a 2D flake of α-MoO$_3$/Au. From those calculations, we remark type I (type II) hyperbolic response with positive (negative) real part of the in-plane permittivities in the α-MoO₃ flake. The RB$_1$ and RB$_4$ present type I $\omega$–$q$ for both crystalline directions. Type II $\omega$–$q$'s are found in the RB$_2$ along [001] and in the RB$_3$ along [100].

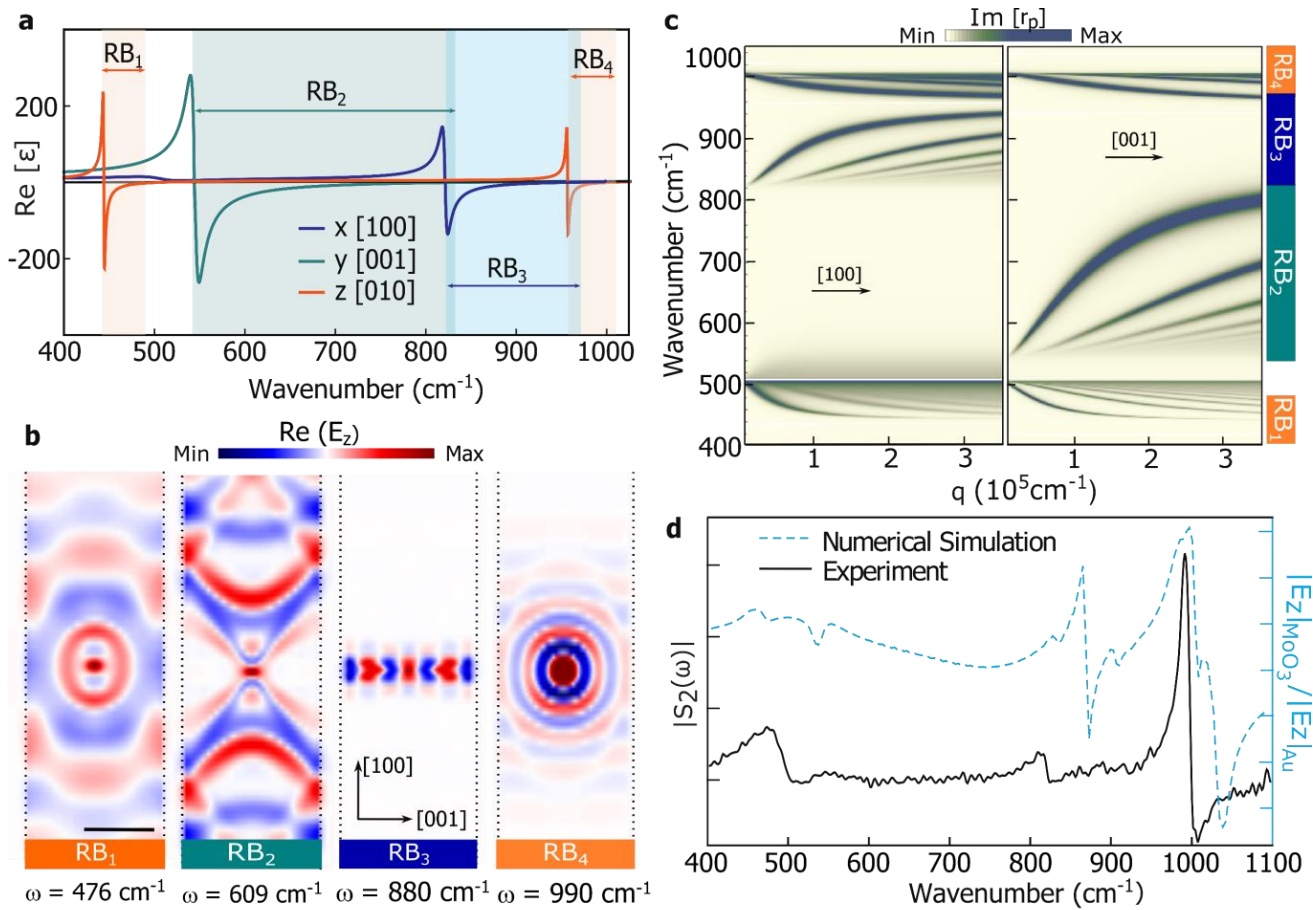

**Figure 2: Polaritonic response of α-MoO₃|** (a) Spectral dependence of the real part of $\varepsilon_{xx}$, $\varepsilon_{yy}$ and $\varepsilon_{zz}$ of an α-MoO₃ crystal. The x, y and z axes are coincident with the crystallographic axes [100], [001] and [010], respectively. Shaded regions indicate the four Reststrahlen bands (RB$_1$, RB$_2$, RB$_3$, and RB$_4$). (b) Simulated Re($E_z$) field distributions of polaritons on a α-MoO₃ nanobelt of w = 1 μm and t = 100 nm on top of gold for: $\omega$ = 476 cm$^{-1}$ in RB$_1$, $\omega$ = 609 cm$^{-1}$ in RB$_2$, $\omega$ = 880 cm$^{-1}$ in RB$_3$ and $\omega$ = 990 cm$^{-1}$ in RB$_4$. The scale bar represents 500nm. (c) Calculated false-color map of the imaginary part of $r_p(\omega, q)$, Im $[r_p(\omega, q)]$, for the air/(2D)α-MoO₃/Au structure along the [100] and [001] crystallographic directions. (d) Normalized SINS amplitude and corresponding simulation spectra of α-MoO₃ nanobelt (w = 1 μm and t = 100 nm) on Au.

To verify how such anisotropy influences the polaritonic response, we analyze by SINS (Figure 3a) different α-MoO$_3$ nanobelts. The SINS point spectrum, in Figure 2e, taken in the center of an isolated nanobelt with w = 490 nm and t = 107 nm on a Au substrate, reveals optical phonon resonances[18]: the stronger peak at 989 cm$^{-1}$ is attributed to stretching mode of Mo-terminal oxygen (O$_1$) with an indicator of the layered orthorhombic MoO$_3$ phase; and two smaller ones at 810 and 475 cm$^{-1}$ ascribed to the stretching mode of oxygen, in Mo–O–Mo bonds, O$_3$ and O$_2$ atoms linked to two or three molybdenum atoms, respectively[18]. These phononic resonances are reproduced in the simulated point-spectrum for an analogous nanobelt (see Methods for more details). We note the overall correspondence in lineshape with the main phonon resonances appearing at approximate spectral positions either in the experiment and the modeling. Small divergences can be attributed to (i) shape inhomogeneities of the real materials (like the rough Au surface) compared with the idealized rectangular nanobelt and fully plane Au surface used in the modeling and (ii) differences between the theoretical and experimental values of permittivity of α-MoO$_3$. Such good agreement between experiment and simulations confirms that our simulation results provide good support to interpret the experimental data in the following.

Complementary, we employed IR SINS (Figure 3a) and pseudo-heterodyne (PH) s-SNOM nanoimaging (Figure 3b) to experimentally access the real-space visualization of HP$^2$ in α-MoO$_3$ nanobelts. In both techniques, the IR radiation is strongly confined at the apex of a metallic AFM tip allowing the launching and detection of HP$^2$s waves due to the momentum match between the nanoprobe and these quasiparticles. With the IR SINS experiment, one can access the broadband response of the nanobelt in a single-shot-like fashion across the mid to far-IR ranges at once. In this case, hyperspectral imaging are necessary for a narrowband analysis (Figure 3c,d). Yet the PH s-SNOM, despite it is not possible to access the full spectral response of SINS, the modality allows fast and highly sensitive narrowband nanoimaging at selected frequencies (Fig. 3e). Polariton fringes at different frequencies over the RB$_1$ an RB$_4$, can be visualized in Figure 3c from nanoimaging reconstructed from SINS hyperspectral measurement on a α-MoO3 nanobelt (w = 667nm, t = 129nm). The fringes result from the interference among type I polaritons waves travelling inside the crystal with strong dependence on the excitation frequency. Similar measurements at RB$_4$ and RB$_3$ for an α-MoO$_3$ nanobelt (w = 1.3 µm, t = 149 nm) are presented in Figure 3d. Additionally, using a tunable mid-IR s-SNOM, we present in Figure 3e near-field images of α-MoO$_3$ nanobelt with $w$ = 590 nm, $t$ = 113 nm taken at frequencies residing inside the RB$_4$.

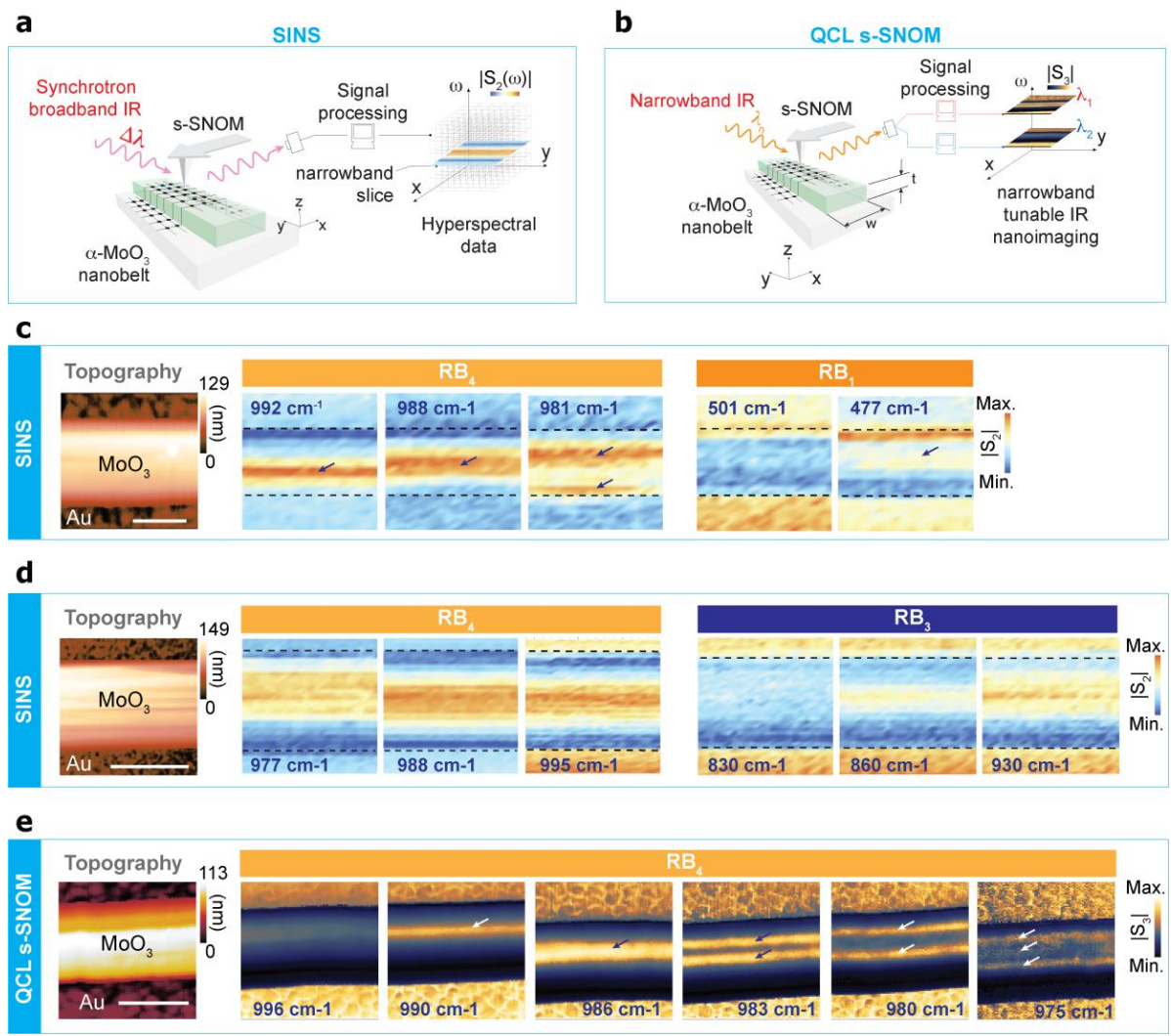

**Figure 3| Far- and mid-IR real-space imaging of HP$^2$ modes in α-MoO$_3$ nanobelt|** Schematics of the (a) SINS experiment for ultrabroadband hyperspectral imaging and (b) QCL s-SNOM experiment for narrowband imaging. (c, d) AFM topography and SINS amplitude ($|S_2|$) narrowband maps reconstructed from a hyperspectral image of α-MoO$_3$

nanobelts/Au in all RBs. Nanobelts dimensions in (c) $w$ = 667 nm, $t$ = 129 nm and, (d), $w$ = 1.3 µm, $t$ = 149 nm. The horizontal, blue-dashed lines indicate the nanobelts edges in correspondence with the topographies. (e) AFM topography and s-SNOM $|S_3|$ on a α-MoO$_3$ nanobelt/Au ($w$ = 590 nm, $t$ = 113 nm) for different $\omega$s in the RB$_4$. Scale bars: 400 nm in (c), 1 µm in (d) and 500nm in (e).

The experimental data (Figures 3c-e) demonstrate spectral bands exhibiting a series of amplitude maxima along the [100] crystal direction (indicated by the arrows). The mechanism that regulates the waves patterns observed in Figures 3c-e is described by tip-launched HP$^2$ waves, which travels across the x-axis of the nanobelts and are reflected by the side edges acquired a round-trip phase. The overlapping of all leads to the formation of standing waves, a FP cavity. The number of maxima ($n$ is the FP resonance order) is related to the excitation $\omega$ within the RBs, indicating a clear dependence of the near-field distribution with the polariton dispersion. Due to the asymmetrical shape of the nanobelts, as seen from AFM topography, the near-field peaks are dislocated from the nanobelt center. Yet, we observe the formation of higher $n$, especially in Figure 3e, in which the s-SNOM nanoimaging presents higher optical contrast due to the superior output power of the laser in comparison to the synchrotron in narrowband experiments.

For a comprehensive assessment of the phenomena observed and as supplementary evidence of nanocavity modes in α-MoO$_3$ nanobelts, we show in Figure 4a the simulated (see Methods for numerical simulation details) and experimental SINS spectral linescan of w = 490 nm and t = 107 nm nanobelt fully covering all RBs. From the experiment, we note the formation of different standing waves patterns along the crystalline axis [100] for each RB due to multiple reflection of HP$^2$ waves at the nanobelt edges. In both simulated and experimental results, the polaritonic activity is not observed in the [001] direction since the measurements are performed far from the extremities of the nanobelt. Such close agreements between experiments and simulations confirm the nanobelt crystalline orientation with respect to the substrate plane and give valuable insights into the HP$^2$ modes ruled by the cavity geometry formed between the nanobelt edges.

In Figure 4b, we zoom in the RB$_1$ (440-500 cm$^{-1}$), RB$_3$ (821-963 cm$^{-1}$) and RB$_4$ (957-1007 cm$^{-1}$) from the experimental linescan in Figure 4a with HP$^2$ profiles extracted at selected frequencies. Standing waves patterns are noteworthy with well-defined number of amplitude maxima, mainly in the RB$_3$ and RB$_4$, indicating different resonant orders of a FP cavity. As a consequence of the different hyperbolicities of each band (Figure 2c), in type I band, the distance between maxima of the spectral linescan increases as we can see in RB4 band. However, in type II band, the opposite trend happens was observed in the RB3 band. Inside the RB1, although standing wave patterns were not observed, we detected polariton activity within the band.

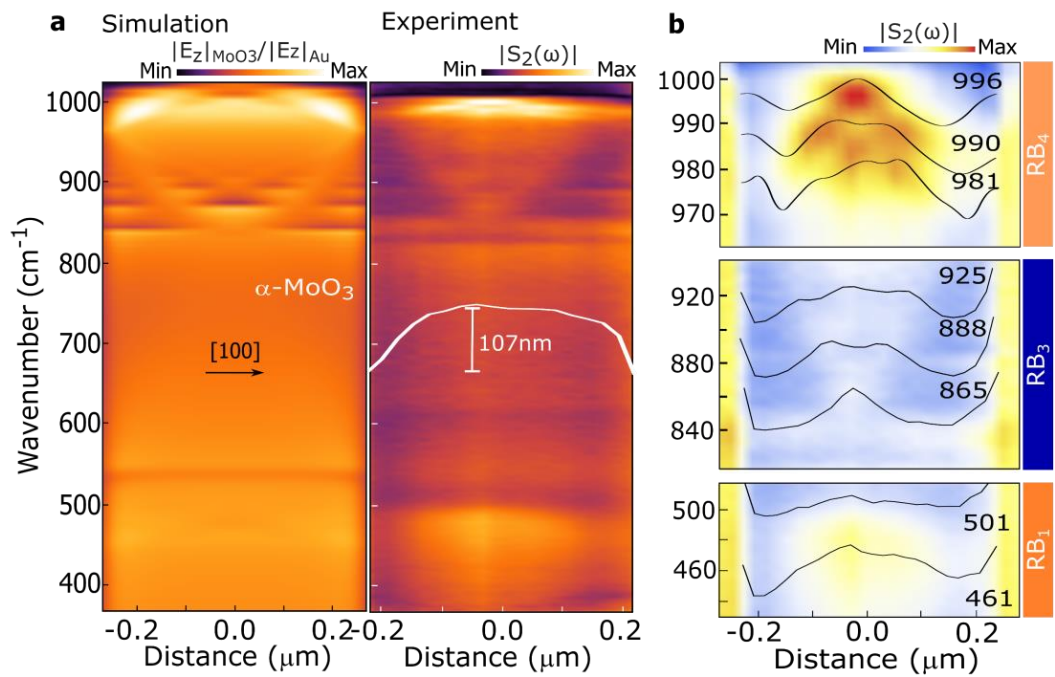

**Figure 4| SINS spectral linescan of α-MoO$_3$ nanobelt with w = 490 nm and t = 107 nm|** (a) Simulated and Experimental spectral linescan and AFM line profile in nanobelt. The RB$_1$, RB$_3$ and RB$_4$ predicted by the theory along the [100] direction are observed. (b) Magnified portions from the experimental linescan in (a) highlighting RB$_4$, RB$_3$ and RB$_1$ from which profiles (black curves) for selected ωs were extracted. For each peak *n* we observed oscillations (*n* = 1, 2, and 3) yielding the Fabry–Perot resonances.

In Figure 5a, as a supplementary evidence for the formation of FP resonances in α-MoO$_3$ nanobelts, we show a simulated spectrum obtained for a nanobelt with w = 1.3 µm and t =267nm and, therefore, for different cavity parameters (See section S4 in the supplementary information for the comparison with experimental result). Figure 5b presents the amplitude profile for selected frequencies inside all RBs shown in Figure 5a. The different resonant FP orders *n* are clearly distinguishable, occurring within all RBs, except for RB$_2$ for the reasons already mentioned. The cavity modes are qualitatively explained in terms of FP resonances, where M$_0$ mode (defined as the mode accessed experimentally in this work - see supporting information) travel across the nanobelt and reflects at the side edges, giving rise to standing waves patterns. The polariton M$_0$ wavelength ($\lambda_{M0}$) for each resonant frequency were calculated, in good approximation, through $\lambda_{M0} = \frac{\delta}{2}$, where $\delta$ is the distance between adjacent near-field maxima at the center of the nanobelt[35]. The extracted correspondent momentum values $(q)$ for the frequencies are plotted (black circles) in Figure 5c, along the (ω-q) dispersion relation for the M$_0$ modes, calculated for an infinite α-MoO$_3$ 2D slab of thickness t = 267nm. We note a reasonable agreement between the theoretical dispersion and the $q$ extracted from the simulations, leading to the conclusion that the FP cavity theory explain the modes transversal to the nanobelt axis.

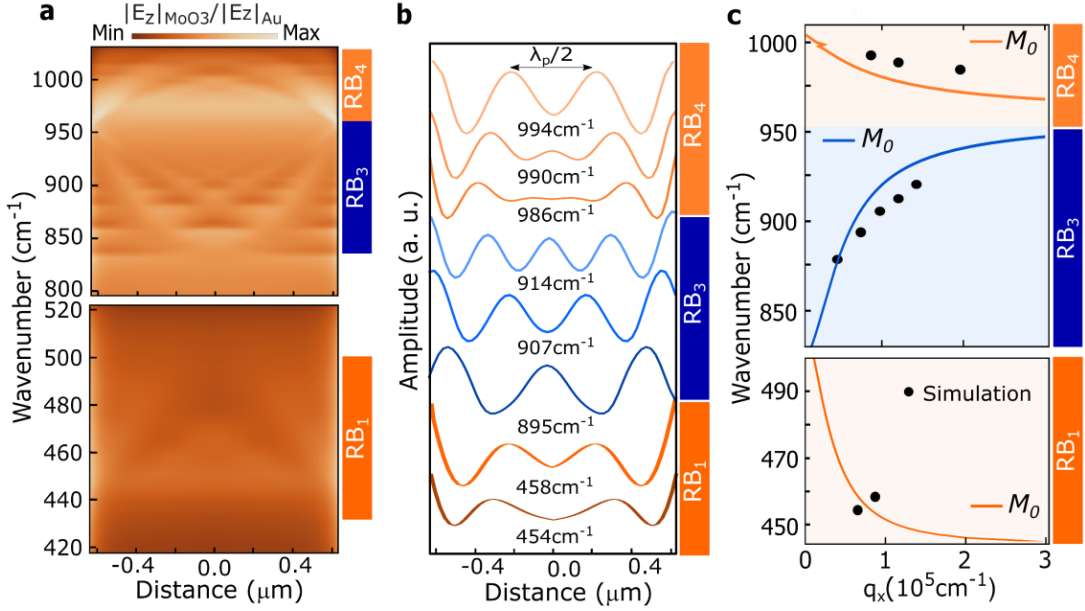

**Figure 5| Simulated visualization of FP cavity modes in α-MoO₃ nanobelt of w = 1.3 μm and t = 267 nm|**(a) Simulated spectral linescan across the nanobelt a top of Au showing the RB$_1$, RB$_3$ and RB$_4$. (b) Polariton profiles for selected $\omega$s acquired in (a). (c) Dispersion curve of HP² modes $M_0$ calculated analytically, see supporting information, (solid curve) at mid to far wavelengths along the [100] direction. Black circles are the values of $q$ of the profiles in (b).

Finally, we estimated a confinement factor ($\lambda_{Ir}/\lambda_{M0}$) of approximately 80 times that of the light in free space ($\lambda_{IR} = 1/\omega$) at $\omega$ = 990 cm$^{-1}$ excitation frequency, which is comparable to ultimate values found in other layered 2D materials[34,45,46]. Therefore, our analysis provides a direct experimental evidence of that α-MoO₃ nanobelts form a Fabry-Perot cavity for HP² modes. We emphasize that whilst type I and II HP² waves have been imaged by s- SNOM using narrow-band lasers in the Mid-IR range[20,22,28] and using free-electron laser[47] in the THz-IR[29], images for type I HP² shown here by SINS in far-IR range are unprecedented.

**Conclusion**

In summary, we used SINS and s-SNOM to map the dispersion of HP² modes in α-MoO₃ nanobelts at mid to far-IR wavelengths. Comparing such nanoimaging with a detailed FDTD-simulation, we conclude that α-MoO₃ nanobelts exhibit Fabry–Perot resonances which are dependent of the sample dimensions.

Additionally, we estimated a confinement factor of about 80 times for the $M_0$ modes. As a further step into applications, we studied the ultrabroadband nanocavity properties of as-grown α-$MoO_3$ nanobelts exploring their lithography-free advantage. For these reasons, we believe that this work offers new prospects to the field of nanophotonics, whereby one may incorporate a single material for resonator applications. The knowledge about the vibrational and optical properties of this low-dimensional material brings insights into new opportunities for studying enhanced light–matter interaction in the mid- to far-IR range.

**Materials and Methods**

*Growth.* The nanobelts of molybdenum trioxide were synthesized by the hydrothermal method proposed by Zhang[12], using Heterocyclic ammonium tetrahydrate CRQ and Nitric Acid ($HNO_3$ 65%) brand Dynamics Química Contemporânea e Ltda. Ammonium heptamolybdate tetrahydrate (AHM) was used as a source of molybdenum. All the chemical reagents were used without further purification, as follows: Initially, 0.4 mmol of AHM was diluted in 20 ml of distilled water. After 30 min of stirring, was obtaining a clear and colorless solution. Then, 10 mL of $HNO_3$ was added and again, the solution has stirred for 30 min. The solution was placed in a sealed autoclave and oven-heated at 180 °C for 20 hours at rate 10 °C / min. After 20 hours, the autoclave was removed from the oven and allowed to cool naturally until reaches room temperature. The product was centrifuged at 5000 rpm for 15 min. After centrifuge time, the solid part remains at the bottom of the tubes, and the supernatant liquid was removed. In this way, the material was washed (repeating the centrifugation procedure and withdrawing the supernatant) for 20 times. The first 10 washes were carried out with distilled water, and the other 10 with ethyl alcohol. The resulting material was oven-dried at 70 °C for 24 hours.

*Scanning electron microscopy* (SEM) images of the nanobelts were performed in performed in a JEOL model JSM6380-LV microscope. The SEM image of the nanobelts shows rectangular cross-section shape, in Figures 2a and 2b with lateral sizes between 50 and 500 nm, and length up to 50 mm.

*Raman.* A table Raman spectrometer was used, Advantage 532 model from SciAps with green laser (532 nm). The synthesized (powdered) material and the commercial molybdenum trioxide (Vitec brand) was characterized. In this way, a comparative analysis can be carried out between the two materials. The material (powder) was placed in cylindrical metal support with a diameter of 6.50 mm and compacted. The spectra were taken with power laser of 0.2 W during 25 s.

*Synchrotron Infrared Nanospectroscopy (SINS).* SINS experiments were performed at the Advanced Light Source (ALS)[30] and at the Brazilian Synchrotron Light Laboratory (LNLS)[31],[32]. Both beamlines use a similar optical setup comprising of an asymmetric Michelson interferometer within a commercial s-SNOM microscope (neaSNOM, Neaspec GmBH). The s-SNOM microscope is an atomic force microscope (AFM) that usually operates in non-contact (tapping) mode with suitable optics to focus and collect the incident and scattered light, respectively. In SINS, broadband IR synchrotron light illuminates the tip oscillating at its fundamental mechanical frequency Ω (~250 kHz) and induces an effective polarization of the tip that enhances light interaction with the sample. The optical near-field signal ($S_m$), which originates from a sample area comparable to the tip radius (around 25 nm in our case), is extracted from the detected back-scattered light by demodulating the signal at harmonics (m) of Ω, with m ≥ 2 with a lock-in amplifier. In SINS, the backscattered light interferes with a reference IR synchrotron beam from a scanning mirror. The Fourier transform of the resultant interferogram yields the IR amplitude $|S_m(\omega)|$ and phase $\Phi_m(w)$ spectra of the complex optical near-field signal $S_m(\omega) = |S_m(\omega)|e^{i\Phi_m(\omega)}$. All SINS spectra here are measured with second harmonic demodulation, m = 2 to yield $S_2(\omega)$. For the mid-IR measurements, we used a ZnSe beamsplitter and a mercury cadmium telluride detector (MCT, IR Associates) for the interferometric setup, whereas for the far-IR measurements, we used a KRS-5 beamsplitter and a custom, helium-cooled Ge:Cu photoconductor, which provides broadband spectral detection down to 320 cm$^{-1}$.

*Infrared nanoimaging.* The infrared nanoimaging of polaritons in α-MoO$_3$ was performed by s-SNOM (Neasnom, Neaspec GmbH) illuminated by a tunable narrowband mid-infrared quantum cascade laser (Daylight Solutions MirCat) with frequency coverage from 930 to 1730 cm$^{-1}$ (minimum frequency step of 1 cm$^{-1}$), in Figure 3e, and broadband IR synchrotron light coverage 320 to 3000 cm$^{-1}$ (Fourier Transform processing with 10 cm$^{-1}$ spectral resolution), in Figures 3c and 3d. For more detail about SINS data acquisition and data processing (hyperspectral images and spectral linescan) see the Supplementary Information.

*FDTD simulations.* The simulations presented in this paper were performed in the commercial software Lumerical FDTD v8.23. The dielectric function of the gold substrate was obtained from Olman *et al*[48], while the dielectric function of α-MoO$_3$ was calculated through the Lorentz model (See supplementary). The Re(E$_z$) fields in figure 2d were excited by a dipole of height H = 100 nm and monitored by a frequency-power plane monitor located at a height h = 20 nm above the crystal surface. The far-fields of the dipole were filtered by applying a starting apodization located at a timeframe of t = 500 fs with time-width of 200 fs. Standard PML boundaries were utilized to absorb the fields outside the simulation region and the simulations were early terminated at t = 3250 fs to prevent simulation instabilities. The simulated point- and line-spectra presented in figures 2e, 4a, and 5a, were performed by approximating the near-field behavior of the s-SNOM tip to a dipole source[49]. Since the dipole moment of the source is constant, tip-sampling coupling effects are not considered in our model. The simulated point-spectrum was obtained by measuring the $|Ez|_{MoO3}$ with frequency-power point monitor located between the dipole and the crystal surface. These values were normalized by the simulated $|Ez|_{Au}$ fields, calculated on the Au substrate without the crystal. The simulated line-spectra were calculated by obtaining equally spaced point-spectra along the nanobelts x-direction. The height of the monitors, h, and the dipole source, H, in respect to the crystal surface were, respectively: (39 nm, 200 nm) for figures 2e and 4a, and (19 nm, 300 nm).

**Acknowledgements**

All authors thank the Brazilian Synchrotron Light Laboratory (LNLS) and Advanced Light Source (ALS) for providing beamtime for the experiments. NeaSpec GmbH is acknowledged for the technical assistance. Angelo L. Gobbi and Maria H. de Oliveira Piazetta, from the Brazilian Nanotechnology National Laboratory. I.D.B, T.A.C, E.G.O, A.M.B.G, and D.C.B.A acknowledges the financial support from the Brazilian Nanocarbon Institute of Science and Technology (INCT/Nanocarbono). I.D.B. acknowledges the support from CNPq through the research grant 311327/2020-6. R.O.F. and R.A.M. acknowledge the FAPESP project 2019/08818-9. R.O.F. acknowledges the support from CNPq through the research grant 311564/2018-6 and FAPESP Young Investigator grant 2019/14017-9. T.A.C, E.G.O, A.M.B.G, and D.C.B.A also like to thank the support of the Universidade Federal de Mato Grosso do Sul - UFMS / MEC – Brazil. F.C.B.M. and F.H.F acknowledge the CNPq project 140594/2020-5. The authors acknowledge CNPq for financial support. The Electron Microscopy Center at UFMG and MULTILAM-UFMS is also greatly acknowledged. This study was financed in part by the Coordenação de Aperfeiçoamento de Pessoal de Nível Superior – Brasil (CAPES) – Finance Code 001. This research used resources of the Advanced Light Source, a U.S. DOE Office of Science User Facility under contract no. DE-AC02-05CH11231.

**Competing interests**

The authors declare no competing interests.

# Ultrabroadband nanocavity of hyperbolic phonon polaritons in 1D-like α-MoO₃

Ingrid D. Barcelos[1†*], Thalita A. Canassa[2*], Rafael A. Mayer[1,3†*], Flavio H. Feres[1,3], Eynara G. de Oliveira[4], Alem-Mar B. Goncalves[2,4], Hans A. Bechtel[5], Raul O. Freitas[1], Francisco C. B. Maia[1], Diego C. B. Alves[2,4]

[1] *Brazilian Synchrotron Light Laboratory (LNLS), Brazilian Center for Research in Energy and Materials (CNPEM), Zip Code 13083-970, Campinas, Sao Paulo, Brazil.*

[2] *Instituto de Química, Universidade Federal de Mato Grosso do Sul (UFMS), Zip Code 79070-900, Campo Grande, Mato Grosso do Sul, Brazil.*

[3] *Physics Department, Gleb Wataghin Physics Institute, University of Campinas (Unicamp), 13083-859 Campinas, Sao Paulo, Brazil*

[4] *Instituto de Física, Universidade Federal de Mato Grosso do Sul (UFMS), Zip Code 79070-900, Campo Grande, Mato Grosso do Sul, Brazil.*

[5] *Advanced Light Source, Lawrence Berkeley National Laboratory, Berkeley, California 94720, USA.*

*These authors contributed equally to this work.
†Corresponding-Author: ingrid.barcelos@lnls.br; rafael.mayer@lnls.br

**Section 1 (S1): SINS broadband imaging of α-MoO₃**

Complementarily, SINS broadband imaging unveils morphology (AFM topography) and broadband local reflectivity of α-MoO$_3$ nanobelt transferred to Au substrate, a standard configuration for all samples analyzed in this work. The AFM topography corroborates the SEM morphological analysis while the s-SNOM amplitude maps highlight broadband reflectivity indicating a preliminary and qualitative view of optical confinement in the nanobelts. |S$_2$| and |S$_3$| represent the 2$^{nd}$ and 3$^{rd}$ harmonics of the s-SNOM tip demodulation, respectively, and confirm the high signal-to-noise ratio and background free quality of the analysis.

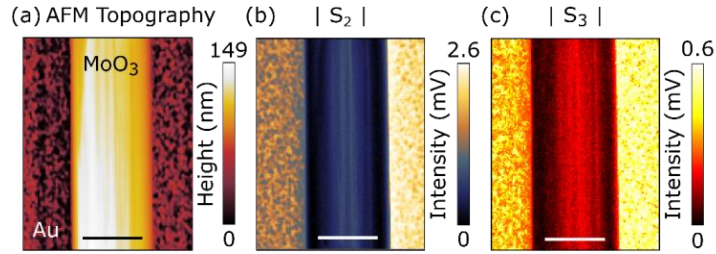

Figure S1: (a) Morphology (AFM topography) and broadband reflectivity, (b) |S$_2$| and (c) |S$_3$|, nanoscale images of an isolated α-MoO$_3$ nanobelt/Au simultaneously measured by SINS. All scale bars represent 1 µm.

**Section 2 (S2): Dielectric function of α-MoO$_3$**

In the infrared region, the optical response of α-MoO$_3$ is dominated by the phonon absorption. The components of the tensor $\overleftrightarrow{\varepsilon}$ are defined in relation to the uniaxial optical axis of the crystal (z-axis).

$$\overleftrightarrow{\varepsilon} = \begin{pmatrix} \varepsilon_{xx} & 0 & 0 \\ 0 & \varepsilon_{yy} & 0 \\ 0 & 0 & \varepsilon_{zz} \end{pmatrix}; \; \varepsilon_{xx} \neq \varepsilon_{yy} \neq \qquad \qquad 1$$

in which the in-plane and out-of-plane components are defined as: orthogonal to the *z-axis* $\varepsilon_{xx} = \varepsilon_{yy}$ and parallel to the *z*-axis $\varepsilon_{zz}$.

The dielectric function, ε, can be described by a Lorentz oscillator model according to

$$\varepsilon_a = \varepsilon_{a,\infty}\left(1 + \sum_j \frac{\left(\omega_{LO,j}^a\right)^2 - \left(\omega_{TO,j}^a\right)^2}{\left(\omega_{TO,j}^a\right)^2 - \omega^2 - i\omega\Gamma_j}\right), \qquad 2$$

where $a$ denotes the $\overleftrightarrow{\varepsilon}$ component, with $a = xx, yy$ or $zz$, having a number $j$ of active optical phonons. $\varepsilon_{a,\infty}$ is the high-frequency term, $\omega$ is the excitation frequency and $\Gamma_j$, the dielectric damping factor, extracted from experimental results in Ref. [1,2].

| Main axis $a$ | Mode index $j$ | $\varepsilon_{x,\infty}$ | $\omega_{TO}$ (cm$^{-1}$) | $\omega_{LO}$ (cm$^{-1}$) | $\Gamma$ (cm$^{-1}$) |
|---|---|---|---|---|---|
| x | 1 | 5.86 | 506.7 | 534.3 | 49.1 |
| x | 2 | | 821.4 | 963 | 6 |

| | | $\varepsilon_{x,\infty}$ | $\omega_{TO}$ (cm$^{-1}$) | $\omega_{LO}$ (cm$^{-1}$) | $\Gamma$ (cm$^{-1}$) |
|---|---|---|---|---|---|
| x | 3 | | 998.7 | 999.2 | 0.35 |
| | | $\varepsilon_{y,\infty}$ | $\omega_{TO}$ (cm$^{-1}$) | $\omega_{LO}$ (cm$^{-1}$) | $\Gamma$ (cm$^{-1}$) |
| y | 1 | 6.59 | 544.6 | 850.1 | 9.5 |
| | | $\varepsilon_{z,\infty}$ | $\omega_{TO}$ (cm$^{-1}$) | $\omega_{LO}$ (cm$^{-1}$) | $\Gamma$ (cm$^{-1}$) |
| z | 1 | 4.47 | 444 | 508 | 1.5 |
| z | 2 | | 956.7 | 1006.9 | 1.5 |

**Table S1.** The experimental value of the parameters used in calculating permittivity.

Using the parameters of the normal oscillation modes, from table S1, we obtained the real part of the components of electrical permittivity shown in Figure 3a in the main text.

The dispersion relation of the α-MoO$_3$ can be described by reflectivity coefficient $r_p(\omega, q)$, considering a multilayer system comprising the Au substrate, α-MoO$_3$ hyperbolic medium and air. The false-color maps, shown in Figure 2c, feature the imaginary part of the Fresnel through the following equations:

$$r_p = \frac{r_1 - r_3 e^{i2k_2^b t}}{1 + r_1 r_3 e^{i2k_2^b t}}, \qquad (3)$$

Where $t$ represents the thickness of the α-MoO$_3$ slab, $r_a = \frac{\varepsilon_\perp k_1^b - \varepsilon_1 k_2^b}{\varepsilon_\perp k_1^b + \varepsilon_1 k_2^b}$, and $r_s = \frac{\varepsilon_3 k_2^b - \varepsilon_1 k_3^b}{\varepsilon_3 k_2^b + \varepsilon_1 k_3^b}$ represent the reflectivity coefficients at the interfaces Air and Au substrate, respectively. The subscripts 1, 2 and 3 denotes air, α-MoO$_3$ and Au. $\varepsilon_1$, $\varepsilon_3$ and $\varepsilon_\perp$ are the relative permittivities of air, Au and components perpendicular to the z-axis. In order to calculate the dispersion relation along the other propagation direction, $\varepsilon_\perp$ can be changed by $\varepsilon_x$ or $\varepsilon_y$ in r$_a$.

### Section 3 (S3): SINS data acquisition and data processing

In our SINS experiment, we collected point spectra with 10 cm$^{-1}$ spectral resolution; the spectra acquisition was done by integrating over 2048 points with 20.1 ms integration time per point. All spectrum are normalized by a reference spectrum acquired from a clean gold surface (100 nm thick Au sputtered on a silicon substrate). The images shown in Figure S1were taken with 150 pixels × 200 pixels with an 11ms integration time per pixel. The spectral linescan in Figure 3c was performed along a line 1 μm length segmented in 50 pixels, with 10 cm$^{-1}$ spectral resolution for each spectrum; 1024 acquisition points with 20 ms integration time per point. The hyperspectral images are recorded at each pixel over a 2.5 μm (1.1 μm) × 2.0 µm (1.1 µm) area segmented in 50 (30) × 40 (30) pixels. Each spectrum in the hyperspectral map resulted from 1 (2) averages over the Fourier transform of an interferogram acquired with 600 µm optical path difference length, yielding 16.6 cm$^{-1}$ spectral resolution, with 10 (5) ms integration time per point. The post-processing of the hyperspectral images data produces a narrow-band map shown in Figure S3 obtained by extracting the amplitude.

**Section 4 (S4): Magnification portion of Far- and mid-IR real-space polaritonic imaging in α-MoO$_3$ nanobelt**

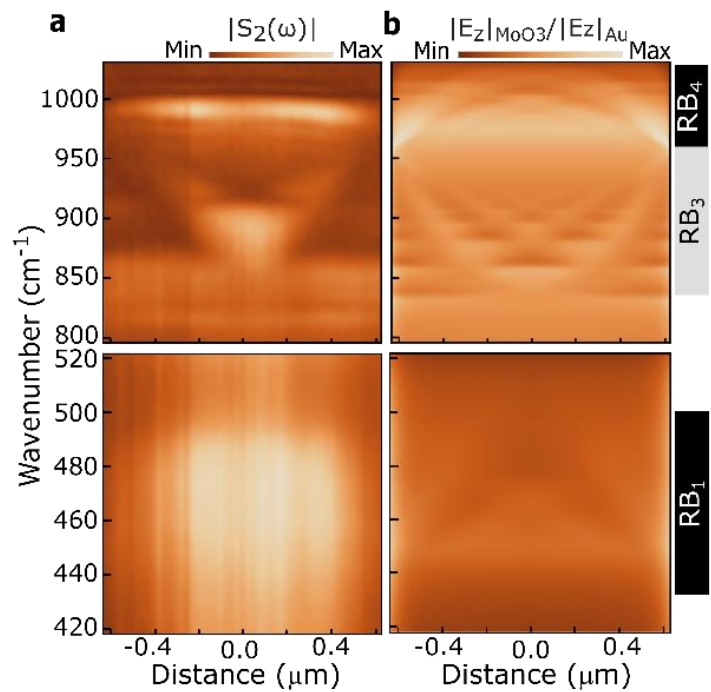

**Figure S2|** **Experimental and Simulated visualization of FP cavity modes in α-MoO$_3$ nanobelt of w = 1.3 µm and t = 267 nm|** (a) Experimental and (b) Simulated spectral linescan across the nanobelt showing the Reststrahlen bands RB$_1$, RB$_3$ and RB$_4$.